\documentclass[a4paper,11pt]{article}
\usepackage{amsmath}
\usepackage{amssymb}
\usepackage[pdftex]{graphicx}
\usepackage{natbib}
\usepackage[pdftex, pdftitle={A Conceivable Origin of Machine Consciousness in the IDLE process},
pdfauthor={Norbert Batfai},
pdfkeywords={Machine consciousness, IDLE process, Minix},
pdfstartview=FitB]{hyperref}
\usepackage{amssymb,amsmath,amsthm,amsfonts}

\newcommand{\keywords}[1]{\par\addvspace\baselineskip\noindent\textbf{Keywords: }\textit{#1}.}

\author{Norbert B\'atfai
\\University of Debrecen
\\Department of Information Technology
\\\texttt{batfai.norbert@inf.unideb.hu}}
\title{A Conceivable Origin of Machine Consciousness in the IDLE process}
\begin{document}
\maketitle

\begin{abstract}
In this short paper, we would like to call professional community's attention to a daring idea that is surely unhelpful, but is exciting for programmers and anyway conflicts with the trend of energy consumption in computer systems.
\end{abstract}

\tableofcontents

\section{Introduction}

Operating systems are very sophisticated programs that can create a dream world for users, in which there are processes, files, windows and etc. In a similar manner, the human consciousness has also created a perfect dream world for us, in which there are smells, colors, words, numbers an so on. However operating systems are still passive programs in this sense that they do exactly what we programmed into them. It is not at all surprising because machines have neither free will nor consciousness today. Namely, in the case of operating systems, the IDLE process will be scheduled if there is nothing to do in the system. In the idle process our computer is doing nothing.

A brief outline of the content of this paper is as follows. In the next section, we introduce the idea of Upright Operating Systems in conceptual level. In section 3, first step will have been taken towards implementation in such a way that we will have a closer look to the replacing IDLE process in Minix operating system. 

\section{Upright Operating Systems}

We set ourselves the aim of making some computing task in the idle process of the kernel. An open source operating system needs to be selected if this goal is to be achieved. We have chosen the Minix3 system \citep{minix3}. The operating systems in which the IDLE process is replaced with some computing task will be called Upright Operating Systems. 

The using of the word "Upright" is an indication of the Upright Man, more specifically of the period of becoming human during which Homo Erectus may have taken an upright position.

\subsection{What would be the cons?}

Why would a computing have to be supported in the kernel space? Why not sufficient a simple, entirely user level server program, which is started by the user, whenever the user wishes to run it? In addition to the exclusion of the classical IDLE process goes against the trend of energy consumption in computer systems.

\subsection{What would be the pros?}

We have only one argument that is principled or even also ethical: this would not be a stoppable program. We will not be able to stop the computing whenever we decide to use an Upright Operating Systems.  

\subsection{What should we compute?}

What computing should be taken at kernel level? Essentially, two approaches may be envisaged. In the first approach, the operating system would observe itself operating, in such a way that it would make notes during its operation, which notes will be processed in idle periods. For example, in the case of Minix, the IPC traffic may be observed by the Minix kernel \citep[examples on pp. 219]{minix3-book}, which in turn enabled the kernel to modify settings of a possible SJF (Shortest Job First) or a preemptive SJF (Shortest Remaining Time First) scheduler.  
   
In the other approach, the operating system, for example, with maintaining a common AIML (Artificial Intelligence Markup Language, \citep[chap. 13]{turingtest}) file, would converge toward human consciousness. But we should remark that the most ideas arisen in this approach are also implementable in the user level and are typically distributed.  

\subsubsection{Time delayed systems}

It is an interesting question, how can Libet and Kornhuber's results on the timing of consciousness \citep{libet}, \citep{kornhuber} be implemented into an upright operating system?

\section{Implementations}

The first step towards implementation is to replace the IDLE process. In the case of choosing Minix operating system, it is easy to find the source code which implements IDLE process \citep{minix3-book}. It can be found in file {\tt kernel/arch/i386/klib386.s}. Here the IDLE process is not only a simple infinity loop, but which contains HLT statement which reduces the CPU energy consumption.   

We don't have to do nothing else than to replace the {\tt idle\_task} reference with an own one in the system image table in source {\tt kernel/table.c}. Our simple own "Hello, World!" style {\tt uos\_task} that demonstrates replacing {\tt idle\_task} defined in source {\tt kernel/uos.c} is the following.

\begin{verbatim}
#include "kernel.h"
#include "../lib/other/random.c"
/* ==================================================== *
 *                       uos_task                       *
 *===================================================== */
PUBLIC void uos_task()
{
  long l = LONG_MIN;
  srandom(get_uptime());

  for (;;) {
    if (random() < l++) {
      kprintf("Hello, Vilag!\n");
      l = LONG_MIN;
    }
  }
}
\end{verbatim}

\section{Conclusion and further work}

In the near future, we are going to accomplish the same firs step described above in Linux kernel. For the time being, we are thinking about the question: what computing should be taken at kernel level?



\bibliography{ConceivableOrigin}

\end{document}